\documentclass[12pt]{article}

\usepackage{booktabs}
\usepackage{scicite}
\usepackage{subcaption}
\usepackage{graphicx}
\usepackage{amsmath}
\usepackage{amssymb}
\usepackage{xcolor}

\usepackage{amsthm}
\newtheorem{definition}{Definition}
\newtheorem{prop}{Proposition}
\newtheorem{theorem}{Theorem}

\usepackage{times}

\newenvironment{sciabstract}{%
\begin{quote} \bf}
{\end{quote}}

\title{Capturing Tie Strength with Algebraic Topology}

\author
{Arnab Sarker,$^{1\ast}$ Jean-Baptiste Seby,$^{1}$ Austin R. Benson,$^{2}$ Ali Jadbabaie$^{1}$\\
\\
\normalsize{$^{1}$Institute for Data, Systems, and Society},\\ \normalsize{Massachusetts Institute of Technology, Cambridge, MA 02139, USA.}\\
\normalsize{$^{2}$Department of Computer Science, Cornell University, Ithaca, NY 14850.}
\\
\normalsize{$^\ast$To whom correspondence should be addressed; E-mail:  arnabs@mit.edu.}
}

\date{}

\renewcommand{\hat}{\widehat}

\begin{document} 


\baselineskip24pt


\maketitle


\begin{sciabstract}
   The association between tie strength and social structure is a fundamental topic in the social sciences. 
   We study this association by analyzing tie strength in higher-order networks, an increasingly relevant model which can encode group interactions between three or more individuals. 
   First, we introduce three measures based on algebraic topology which characterize the network context and influence of an edge.
   Our experimental results across 15 datasets indicate that these measures outperform standard network proxies in estimating tie strength. 
   We further find that these measures can replicate and explain a puzzle wherein certain bridging ties are surprisingly strong.
   We then consider a single centrality measure which combines the three initial measures, is highly inversely related to tie strength, and can be interpreted through an information exchange process which highlights ties that have access to useful information. 
   In this sense, we are able to illuminate the information advantages of weak ties due to their network position.
\end{sciabstract}

% Teaser: Algebraic topological tools clarify recent puzzles about tie strength and highlight edges poised to carry valuable information.

% In setting up this template for *Science* papers, we've used both
% the \section command and the \paragraph* command for topical
% divisions.  Which you use will of course depend on the type of paper
% you're writing.  Review Articles tend to have displayed headings, for
% which \section is more appropriate; Research Articles, when they have
% formal topical divisions at all, tend to signal them with bold text
% that runs into the paragraph, for which \paragraph* is the right
% choice.  Either way, use the asterisk (*) modifier, as shown, to

\section{Introduction}

The association between the strength of a tie and that tie's position in the broader social structure has been the focus of a vast literature across the social sciences \cite{aral2022what,aral2011diversity,burt2002social,gilbert2009predicting,granovetter1973strength,mattie2018understanding}.
Tie strength, which captures the intensity of a relationship between two individuals and can include dimensions such as frequency of interaction, intimacy, emotional intensity, and reciprocity~\cite{granovetter1973strength}, has been shown at length to impact substantive outcomes such as job outcomes \cite{granovetter1973strength,granovetter2018getting,kim2017strength}, creativity \cite{uzzi2005collaboration}, political success \cite{padgett1993robust}, and knowledge transfer in organizations \cite{hansen1999search,reagans2003network}.
Further, tie strength is an important feature for data-driven tasks such as link prediction \cite{de2011supervised,li2013modeling,lu2010link} and recommendation in online social networks \cite{rajkumar2022causal,wang2016social}.

At the core of many arguments surrounding the importance of tie strength is that, although tie strength is locally defined between two individuals, this local notion seems to reflect aspects of the tie's position in the broader network topology. 
The relationship between tie strength and network structure was first discussed by Granovetter \cite{granovetter1973strength}, who suggested that bridging ties, i.e. ties between individuals who share no common neighbors, are likely to be weak ties. 
This association between bridging ties and weak ties follows from a premise regarding strong ties and triadic substructures in the social network. 
Namely, if an individual has strong ties to two different alters, then those alters should know one another such that these three individuals should form a triangle in the network.
As a consequence, bridging ties should be weak.
This ultimately leads to the strength of weak ties thesis--- the simple yet powerful notion that weak ties are more likely than strong ties to offer valuable network information.

However, empirical evidence showing that bridging ties are weak has been mixed \cite{aral2022what,kim2023makes,neal2023not,park2018strength}.
In an analysis across 56 datasets, Neal finds that on average, only 51\% of bridges in a dataset can be labeled as weak ties \cite{neal2023not}.
Further, recent results on population-scale networks have found that certain types of bridging ties, particularly those with high tie range,\footnote{Tie range is defined as the second shortest path length associated with an edge \cite{park2018strength}.} can be surprisingly strong \cite{jahani2023long,lyu2022investigating,park2018strength}.
These ``long ties'' with high tie strength can be particularly important for network processes such as information diffusion \cite{park2018strength} and individual outcomes such as economic resilience \cite{jahani2023long}.
Hence, it is particularly important to understand the relationship between tie strength and network structure, with specific emphasis on how different types of bridging ties vary in terms of dyadic strength.

Moreover, while tie strength impacts network structure through the mechanisms Granovetter suggests, network structure can also affect tie strength as shared neighbors can reinforce social ties \cite{gilbert2009predicting,krackhardt2003strength}.
As such, it is important to understand the context in which a tie occurs. 
For example, if two individuals who share a tie commonly interact in the context of larger groups, then these group contexts may shape their relationship (c.f. \cite{collins2004interaction}), whereas if these two individuals commonly interact in one-on-one settings, then tie strength is more likely truly a local function of their relationship (c.f. \cite{burt2024guanxi}).
By identifying the network features which are correlated with tie strength, we may better understand the complex, endogenous relationship between tie strength and network structure.

In this work, we lay a foundation for understanding the association between tie strength and network structure by modeling social structure with a higher-order network. 
Higher-order network models have gained much recent interest in the literature as they can explicitly account for group interactions in which three or more elements of the network interact at once \cite{battiston2020networks,benson2016higher,iacopini2019simplicial,sarker2024higher,schaub2018random,wu2022link}. 
These group interactions, also referred to as higher-order interactions, can display social phenomena which are difficult to capture with traditional models of networks such as social pressure and collective cooperation \cite{alvarez2021evolutionary,asch1955opinions,iacopini2019simplicial}.
Moreover, such higher-order interactions have been theorized to shape broader social outcomes and generate group solidarity, ultimately suggesting that higher-order information is valuable when estimating tie strength \cite{collins2004interaction}.

There are many ways to model a higher-order network, including hypergraphs, bipartite projections, and set systems \cite{battiston2020networks,benson2016higher}, and here we use a simplicial complex to encode the presence of higher-order interactions in our network datasets.
Simplicial complexes have been studied for decades by mathematicians to develop a rich theory in algebraic topology \cite{hatcher2002algebraic, hodge1989theory}, and have been increasingly used in applied domains \cite{benson2016higher,schaub2018random,giusti2016two,tahbaz2010distributed}.
Using a simplicial complex to represent higher-order network data allows us to introduce new structural measures to understand the network position of an edge, and empirically compare these measures to tie strength in social networks to better understand the relationship between network structure and tie strength.

\paragraph{Summary of Results}

We first introduce three structural network measures which use algebraic topology to encode an edge's position within the social structure.
Using the concept of Hodge Decomposition \cite{hatcher2002algebraic}, which states that any flow in a simplicial complex can be decomposed into a gradient component, a curl component, and a harmonic component, we develop three respective measures for each edge in a simplicial complex.
Namely, our method applies the Hodge Decomposition to the indicator flow on an edge to describe the gradient, curl, and harmonic component of an edge.
Empirically, we perform experiments across 15 large-scale datasets which indicate that these measures outperform standard network baselines in estimating (out-of-sample) tie strength.

Theoretically, we show that the gradient component of an edge represents the edge's ability to disconnect the graph, the curl component represents an edge's proximity to higher-order interactions, and the harmonic component represents an edge's closeness to topological obstructions (holes) in the network (c.f. \cite{ghrist2005coverage,hatcher2002algebraic,lim2015hodge,schaub2018random,tahbaz2010distributed}).
Our characterizations further reveal that the gradient and harmonic components can distinguish between different types of bridging ties--- bridges which are ``long ties'' for which the second shortest path length is particularly high tend to have a larger gradient component whereas bridges that span a medium network distance tend to be associated with a larger harmonic component.

These theoretical results, in tandem with the measures' ability to empirically estimate tie strength well, allow us to replicate and explain an apparent puzzle in the network sociology literature regarding the strength of long-range ties. 
Several population-scale datasets show that long ties can be nearly as strong as ties which share common neighbors, and describe a ``U''-shaped relationship between tie range and tie strength \cite{lyu2022investigating,park2018strength}.
We show that the network measures mentioned above recover this ``U''-shape.
We then show how each of the gradient, curl, and harmonic components contribute to the ``U''-shape, offering an explanation for why the shape exists.

In particular, we find that an edge's curl component and its gradient component tend to be positively correlated with tie strength, offering reasons for ties with many common neighbors and ties which span large network distances to be strong, respectively.
This ultimately suggests how network structure is associated with the strength of a tie in a context-dependent manner--- whereas the strength of ties with a high curl component is associated with the presence of social support due to group interactions, the strength of ties with a high gradient component is associated with the apparent dissimilarities between individuals.

Importantly, although the result on the strength of long-range ties appears to be in conflict with Granovetter's original intuition, here we show that the two can be reconciled.
Specifically, we analyze a singular measure, \emph{Edge PageRank}, introduced by Schaub \textit{et al.} \cite{schaub2018random}.
Our experiments show that Edge PageRank is effective in estimating tie strength and is highly inversely correlated with tie strength, i.e. the measure identifies weak ties.
Theoretically, we show that the measure can be written as a combination of the gradient, curl, and harmonic components above, and has a non-monotonic relationship with tie strength such that it can also recover the ``U''-shape relationship between tie strength and tie range.

Although the Edge PageRank measure was initially introduced as a mathematical extension of the classical PageRank measure \cite{page1999pagerank}, here we show that the measure can be interpreted through an interpretable social process by generalizing the frameworks set forth by Bonacich \cite{bonacich1987power} and Friedkin \cite{friedkin1990social,friedkin1991theoretical}.
We show how the Edge PageRank measure can also be interpreted as the outcome of a random information exchange process, which ultimately highlights edges which are in a structural position to transfer useful information.
In this sense, we see that Granovetter's original intuition still holds: our interpretation of the Edge PageRank measure and experiments on tie strength reveal that that weak ties are often in a good structural position to transfer useful information.
However, because Edge PageRank does not emphasize long ties, this suggests an amendment to Granovetter's theory in that while long ties may provide novel information, the utility of this information may decrease as tie range increases (c.f. \cite{kim2023makes}).

Ultimately, these results suggest the importance of incorporating higher-order interactions in social network analysis more broadly, as this additional data captures distinct sociological insights compared to traditional models.

\section{Generating Topological Features with Hodge Decomposition}
In order to describe how we construct network measures based on the Hodge Decomposition, we first describe how we model our data using simplicial complexes, a tool from algebraic topology that extends traditional network models to explicitly encode higher-order interactions.
We then describe some preliminaries related to algebraic topology before formally introducing the measures.

\subsection{Simplicial Complexes}
\label{ss:scs}

We first provide a definition of a simplicial complex, and then discuss the interpretation of the data structure.
\clearpage
\begin{definition}[Simplicial Complex \cite{hatcher2002algebraic}]
Given a set of vertices $V$, a simplicial complex $\mathcal{X}$ is a set of subsets of $V$, i.e. $\mathcal{X} \subseteq 2^V$,
such that $\mathcal{X}$ satisfies the following inclusion property:
\begin{equation} \label{eq:inclusion}
    x \in \mathcal{X} \implies \sigma \in \mathcal{X}\,, \qquad \forall \sigma \subseteq x \,.
\end{equation}
\end{definition}

Each element $x$ of the simplicial complex represents some type of (potentially higher-order) relationship between a subset of vertices in $V$.
Conventionally, if $x$ contains $k + 1$ elements of $V$, then $x$ is referred to as a $k$-simplex.
In Table \ref{tab:data}, we refer to $0$-simplices as nodes, $1$-simplices as edges, and $2$-simplices as filled triangles.

The inclusion property of a simplicial complex states that, if a higher-order interaction $x$ is encoded in the simplicial complex $\mathcal{X}$, then all subsets of $x$ are also contained in $\mathcal{X}$.
For example, if there is a filled triangle $\{v_1, v_2, v_3\}$ in $\mathcal{X}$, then we must also have all relevant edges $\{v_1, v_2 \} \in \mathcal{X}$, $\{v_1, v_3 \} \in \mathcal{X}$, and $\{v_2, v_3 \} \in \mathcal{X}$.\footnote{That being said, the converse need not be true--- there may be three edges $\{a, b\} \in \mathcal{X}$, $\{a, c\} \in \mathcal{X}$, and $\{b, c\} \in \mathcal{X}$, but $\{ a, b, c \} \notin \mathcal{X}$. 
The nodes $a, b,$ and $c$ would then be considered an ``unfilled'' triangle in $\mathcal{X}$, since they would be dyadically connected but not have a filled triangle. That is, $a$, $b$, and $c$ may have interacted in pairwise settings but not had an interaction where all three were co-present in the same situation.}
As such, the interpretation of a simplex in a simplicial complex can often be different than that of, e.g., a hyperedge in a hypergraph in that a simplicial complex often encodes ``relationships'' as opposed to explicit ``interactions'' (c.f. \cite{torres2021and}).

Importantly, the inclusion property of a simplicial complex allows us to represent higher-order interactions of size 4 or larger as a collection of filled triangles in each dataset.
Because higher-order interactions of size larger than 4 will induce simplices of size 3 in the simplicial complex, these larger interactions are still encoded as filled triangles in the simplicial complex and each filled triangle will indicate that a set of three nodes has participated in at least one higher-order interaction with one another.

\subsection{Boundary Operators, Hodge Laplacians, and Hodge Decomposition}
\label{ss:hodge}

A second key consequence of the inclusion property is that it allows us to define boundary operators on the simplicial complex.

To define the boundary operators, it will help notationally to write $V$ for the set of nodes in the simplicial complex, $E$ for the set of edges, and $T$ for the set of filled triangles.
Moreover, we will assume that the nodes have some numbering from $1$ to $|V|$.
For the purposes of this work, we consider two boundary operators: $B_1$, which acts as a signed node-edge incidence matrix, and $B_2$, which acts as a signed edge-triangle incidence matrix.

Formally, the boundary operator $B_1 \in \mathbb{R}^{|V| \times |E|}$ is a matrix where rows correspond to nodes and columns correspond to edges.
For each edge $\{i, j\}$, where $i < j$, $B_1[i, \{i, j\}] = +1$ and $B_1[j, \{i, j\}] = -1$.
All other entries of $B_1$ are equal to $0$.
The boundary operator $B_2 \in \mathbb{R}^{|E| \times |T|}$ has a similar definition.
For every triangle $\{i, j, k\}$, where $i < j < k$, we set $B_2[\{i, j\}, \{i, j, k\}] = B_2[\{j, k\}, \{i, j, k\}] = +1$ and $B_2[\{i, k\}, \{i, j, k\}]  = -1$.
All other entries of $B_2$ are similarly set to 0.

For the purposes of this paper, the important qualities of $B_1$ and $B_2$ are that they are sparse, and hence many mathematical operations become computationally feasible, and that $B_1 B_2 = 0$ due to the signs of the entries in each matrix.
This latter quality will be useful later as it enables us to employ the tool of Hodge Decomposition.
Further, these operators allow us to define the Hodge Laplacian.

\begin{definition}[$1$-Hodge Laplacian \cite{schaub2018random}]\label{def:laplacian}
    For a simplicial complex $\mathcal{X}$ with boundary operators $B_1$ and $B_2$, the $1$-Hodge Laplacian is a matrix $L_1 \in \mathbb{R}^{|E| \times |E|}$ which satisfies
    \begin{equation}\label{eq:hodge_laplacian}
        L_1 =B_1^\top B_1 + B_2 B_2^\top \,.
    \end{equation}
\end{definition}

The $1$-Hodge Laplacian discussed by Hodge \cite{hodge1989theory}, who notes the relationship between this Laplacian operator and the homology groups of a simplicial complex.
Moreover, it is worth noting that the $1$-Hodge Laplacian can be generalized to a $k$-Hodge Laplacian which acts on $k$-simplices for different values of $k$, and that the $0$-Hodge Laplacian corresponds to the standard Laplacian operator defined on graphs.

A particularly important consequence of these definitions is that any edge flow can be decomposed into curl, gradient, and harmonic scores using the Hodge Decomposition.
\begin{definition}[Hodge Decomposition]\label{def:hodge}
    For a vector $v \in \mathbb{R}^{|E|}$, the Hodge Decomposition of $v$ is a set of 3 vectors $v^g, v^c$, and $v^h$ such that $v = v^g + v^c + v^h$.
    Specifically, $v^g$ is the projection of $v$ onto $B_1^\top$, $v^c$ is the projection of $v$ onto $B_2$, and $v^h$ is defined as $v - v^g - v^c.$
\end{definition}
The components $v^g$, $v^c$, and $v^h$ are orthogonal, and a rich theory in algebraic topology has been developed which analyzes the mathematical properties of these quantities (c.f. \cite{hatcher2002algebraic,schaub2018random}).
% Importantly, the three components of the Hodge Decomposition satisfy:
% \[
%     v^g \in \im(B_1^\top)\,, \quad v^c \in \im (B_2) \,, \quad \text{ and } \quad v^h \in \ker (L_1) \,.
% \]

% These properties of the Hodge Decomposition help to interpret the three components as a function of tie range. 
% Roughly, because $\im (B_1^\top)$ is the space of cuts in the graph, this suggests that $v^g$ is associated with global bridges in a graph and ties with larger tie range.
% Because $\im(B_2)$ is the space of functions defined on filled triangles, $v^c$ is associated with triangles in the underlying graph, i.e. ties with a tie range of range 2.
% Finally, $\ker (L_1)$ is well known to be associated with cycles that are not filled by triangles in a simplicial complex (i.e., topological holes), which suggests $v^h$ is associated with a tie range of 3 or larger.

\subsection{Structural Measures from Hodge Decomposition}
Using the well-established notion of the Hodge Decomposition, we are able to introduce structural measures for each edge in the network.
Specifically, we consider the indicator vector of an edge, $\delta_e \in \mathbb{R}^{|E|}$, consider how this vector can be decomposed using the Hodge Decomposition, and then summarize each component with the $2$-norm to get a single structural measure for each component.
Formally, we have the following definition.
\begin{definition}[Gradient, Curl, and Harmonic of an Indicator]
\label{def:hodge_measures}
    For a simplicial complex $\mathcal{X}$ and an edge $e \in \mathcal{X}$, let $\delta_e$ be defined as a vector which is $1$ at the index corresponding to edge $e$ and 0 otherwise. Let $\delta_e^g, \delta_e^c,$ and $\delta_e^h$ represent the weighted Hodge Decomposition of $\delta_e$.
    Then, the magnitude of the gradient, curl, and harmonic components of $\delta_e$ are defined:
    \[ I^g_e = \|\delta_e^g \| \,, \quad I^c_e = \|\delta_e^c \|\,, \quad \text{ and } \quad I^h_e = \|\delta_e^h \| \,,\]
    respectively.
\end{definition}
By projecting the indicator of an edge into these three subspaces, we are able to capture an edges position in the broader higher-order network topology.
Moreover, we summarize each measure using the norm $\| \cdot \|$, which we take to be the standard $2$-norm $\|x\| = \sqrt{\sum_i x_i^2}$.
As we will show, these structural measures estimate tie strength well, naturally replicate a phenomena wherein certain bridging ties are strong, and suggest two potential mechanisms by which a tie can be strong.
% Moreover, these measures naturally relate to tie range, ultimately resolving the puzzling observation wherein certain types of bridging ties are surprisingly strong.

\section{Estimating Tie Strength and Replicating the Strength of Long-Range Ties}

With the structural measures defined, we now first present empirical evidence that these measures accurately estimate tie strength.
We then develop interpretations of the measures to better understand how they describe an edge's position in a simplicial complex and to theoretically relate them to tie range.

\begin{table}[t!]
    \centering
    \caption{Data used for Tie Strength Estimation Experiments. For each dataset, we compute the number of individuals in the network (nodes), pairwise relationships (edges), and triadic relationships (filled triangles). We also compute the density of edges, the fraction of edges observed out of all possible edges, as well as the percent of closed triangles which are filled, where a closed triangle is a set of three nodes who share all three possible dyadic relationships but need not have an explicit triadic interaction.}
\footnotesize
    \begin{tabular}{rrrrrr}
    \toprule
                        Dataset &  Nodes &   Edges &          Edge Density & \shortstack[r]{Filled\\Triangles} & \shortstack[r]{Percent of\\ Closed Triangles \\which are Filled }\\
\midrule
           \texttt{bills-house} &  1,471 &  29,959 & $2.77\times 10^{-02}$ &           16,884 &                                      16.04\% \\
          \texttt{bills-senate} &    295 &  10,555 & $2.43\times 10^{-01}$ &           11,460 &                                       7.20\% \\
           \texttt{coauth-dblp} & 81,427 & 170,516 & $5.14\times 10^{-05}$ &           83,599 &                                      88.91\% \\
           \texttt{college-msg} &  1,899 &  13,838 & $7.68\times 10^{-03}$ &            5,403 &                                      37.73\% \\
   \texttt{contact-high-school} &    327 &   5,818 & $1.09\times 10^{-01}$ &            2,370 &                                       6.93\% \\
      \texttt{contact-hospital} &     73 &   1,340 & $5.10\times 10^{-01}$ &            3,839 &                                      27.70\% \\
\texttt{contact-primary-school} &    242 &   8,317 & $2.85\times 10^{-01}$ &            5,139 &                                       4.95\% \\
    \texttt{contact-university} &    692 &  79,530 & $3.33\times 10^{-01}$ &          436,298 &                                      11.28\% \\
  \texttt{contact-workplace-13} &     95 &   3,151 & $7.06\times 10^{-01}$ &            5,901 &                                      10.52\% \\
  \texttt{contact-workplace-15} &    219 &  11,772 & $4.93\times 10^{-01}$ &           20,367 &                                       7.42\% \\
           \texttt{email-Enron} &    144 &   1,344 & $1.31\times 10^{-01}$ &            1,159 &                                      27.70\% \\
              \texttt{email-Eu} &    986 &  16,064 & $3.31\times 10^{-02}$ &           27,655 &                                      26.22\% \\
        \texttt{india-villages} & 69,217 & 282,787 & $1.18\times 10^{-04}$ &          342,945 &                                      81.14\% \\
                 \texttt{sms-a} & 30,278 &  42,882 & $9.36\times 10^{-05}$ &            1,581 &                                      16.43\% \\
                 \texttt{sms-c} & 11,714 &  17,050 & $2.49\times 10^{-04}$ &            1,114 &                                      20.23\% \\
\bottomrule
    \end{tabular}
    \label{tab:data}
\end{table}

\subsection{Empirical Tie Strength Estimation}

In this section, we establish the empirical use of the structural measures above by establishing a relationship between the network measures above and measures of tie strength.
We find that the Hodge Decomposition measures both outperform network baselines in measuring tie strength and also replicate a puzzle wherein long-range ties can be surprisingly strong.

\subsubsection{Data} 
We perform large-scale data analyses on 15 datasets from diverse domains as described in Table \ref{tab:data} to understand how the measures in Definition \ref{def:hodge_measures} relate to tie strength.
For each dataset, we form a simplicial complex $\mathcal{X}$ for which each edge also has an auxiliary measure of tie strength.
These networks are described as follows:
\begin{itemize}
    \item \texttt{bills-house}, \texttt{bills-senate}. Nodes are members of the United States Congress, and simplices form when individuals co-sponsor bills with one another \cite{fowler2006connecting,fowler2006legislative}. 
    \item \texttt{coauth-dblp}. Nodes are authors of academic papers, and simplices form when two or more authors co-author a paper with one another \cite{agarwal2016women}.
    \item \texttt{contact-high-school},\texttt{contact-hospital}, \texttt{contact-primary-school}, 
    
    \texttt{contact-university}, \texttt{contact-workplace-13}, \texttt{contact-workplace-15}. Nodes are individuals, and simplices form when individuals are in proximity of one another within a short time period according to Bluetooth sensors \cite{genois2018can,sapiezynski2019interaction}.
    \item \texttt{email-Enron}, \texttt{email-Eu}. Nodes are email addresses and a simplex is formed if individuals send or CC one another on emails and reciprocate within a week;
    \texttt{email-Eu} spans over 2 years of communication at a European research institute~\cite{benson2016higher,paranjape2017motifs}, and \\
    \texttt{email-Enron} spans the lifetime of the American energy company Enron~\cite{benson2018simplicial,klimt2004enron} .
    \item \texttt{india-villages}. Nodes are individuals in villages in India, and edges form between individuals based on survey data~\cite{banerjee2013diffusion}, and simplices correspond to individuals living in the same household.
    \item \texttt{sms-a}, \texttt{sms-c}, \texttt{college-msg}. Nodes are individuals, and a simplex forms between individuals if they all send a message to each other within a week~\cite{panzarasa2009patterns,wu2010evidence}. 
\end{itemize}

For all datasets except \texttt{india-villages}, tie strength is based on the number of times two individuals appeared in some interaction together.
In other words, if two individuals are in contact frequently, then their tie is considered to be stronger. 
Since the distribution of frequency of contacts is heavy-tailed, we use the log of the frequency for tie strength in experiments.
For the \texttt{india-villages} dataset, tie strength is determined on a scale of 1 to 10 based on answers to a questionnaire take by each individual~\cite{banerjee2013diffusion}. For this dataset, there are questions regarding household membership which we omit in tie strength computation as to prevent data leakage, since household membership determines higher-order interactions.

% Table: Tie Strength Prediction Results
\begin{table}[]
    \centering
    \caption{Accuracy of predicting tie strength with different sets of regressors. Each entry is the test accuracy of a linear regression, computed using a 10-fold cross-validation. 
    Bolded entries indicate when a measure statistically significantly outperforms all others.}
    \footnotesize
   \begin{tabular}{rrrrr}
\toprule
 &                          Hodge Components &           \shortstack[r]{Network\\Baseline \cite{mattie2018understanding}} &              \shortstack[r]{Unweighted\\Overlap} &          \shortstack[r]{Node\\PageRank\\Baseline}\\
\midrule
\texttt{bills-house}            &  \textbf{0.351} \scriptsize{($\pm$0.006)} &  0.105 \scriptsize{($\pm$0.003)} &  0.098 \scriptsize{($\pm$0.004)} &      0.067 \scriptsize{($\pm$0.003)} \\
\texttt{bills-senate}           &  \textbf{0.230} \scriptsize{($\pm$0.006)} &  0.185 \scriptsize{($\pm$0.006)} &  0.157 \scriptsize{($\pm$0.005)} &  0.113 \scriptsize{($\pm$0.005)} \\
\texttt{coauth-dblp}            &  \textbf{0.141} \scriptsize{($\pm$0.001)} &  0.093 \scriptsize{($\pm$0.001)} &  0.007 \scriptsize{($\pm$0.000)} &      0.004 \scriptsize{($\pm$0.000)} \\
\texttt{college-msg}            &  \textbf{0.081} \scriptsize{($\pm$0.006)} &  0.025 \scriptsize{($\pm$0.003)} &  0.018 \scriptsize{($\pm$0.002)} &  0.047 \scriptsize{($\pm$0.004)} \\
\texttt{contact-high-school}    &  \textbf{0.403} \scriptsize{($\pm$0.009)} &  0.214 \scriptsize{($\pm$0.006)} &  0.187 \scriptsize{($\pm$0.007)} &  0.009 \scriptsize{($\pm$0.003)} \\
\texttt{contact-hospital}       &           0.168 \scriptsize{($\pm$0.016)} &  0.209 \scriptsize{($\pm$0.018)} &  0.182 \scriptsize{($\pm$0.019)} &  0.113 \scriptsize{($\pm$0.016)} \\
\texttt{contact-primary-school} &  \textbf{0.614} \scriptsize{($\pm$0.005)} &  0.349 \scriptsize{($\pm$0.006)} &  0.309 \scriptsize{($\pm$0.006)} &  0.009 \scriptsize{($\pm$0.003)} \\
\texttt{contact-university}     &           0.183 \scriptsize{($\pm$0.002)} &  0.188 \scriptsize{($\pm$0.003)} &  0.116 \scriptsize{($\pm$0.002)} &      0.089 \scriptsize{($\pm$0.002)} \\
\texttt{contact-workplace-13}   &  \textbf{0.176} \scriptsize{($\pm$0.006)} &  0.138 \scriptsize{($\pm$0.011)} &  0.089 \scriptsize{($\pm$0.008)} &  0.042 \scriptsize{($\pm$0.007)} \\
\texttt{contact-workplace-15}   &  \textbf{0.215} \scriptsize{($\pm$0.004)} &  0.150 \scriptsize{($\pm$0.004)} &  0.133 \scriptsize{($\pm$0.004)} &  0.005 \scriptsize{($\pm$0.001)} \\
\texttt{email-Enron}            &           0.247 \scriptsize{($\pm$0.020)} &  0.202 \scriptsize{($\pm$0.012)} &  0.166 \scriptsize{($\pm$0.012)} &  0.052 \scriptsize{($\pm$0.016)} \\
\texttt{email-Eu}               &           0.157 \scriptsize{($\pm$0.004)} &  0.161 \scriptsize{($\pm$0.006)} &  0.146 \scriptsize{($\pm$0.006)} &  0.010 \scriptsize{($\pm$0.001)} \\
\texttt{india-villages}         &  \textbf{0.671} \scriptsize{($\pm$0.001)} &  0.534 \scriptsize{($\pm$0.001)} &  0.482 \scriptsize{($\pm$0.001)} &  0.251 \scriptsize{($\pm$0.001)} \\
\texttt{sms-a}                  &  \textbf{0.049} \scriptsize{($\pm$0.003)} &  0.012 \scriptsize{($\pm$0.001)} &  0.011 \scriptsize{($\pm$0.001)} &  0.005 \scriptsize{($\pm$0.001)} \\
\texttt{sms-c}                  &  \textbf{0.050} \scriptsize{($\pm$0.003)} &  0.022 \scriptsize{($\pm$0.003)} &  0.014 \scriptsize{($\pm$0.003)} &      0.006 \scriptsize{($\pm$0.002)} \\
\bottomrule
\end{tabular}
\label{tab:hodge_prediction}
\end{table}

\subsubsection{Results} Table \ref{tab:hodge_prediction} summarizes our results on tie strength estimation.
Specifically, we develop a linear model of tie strength using the features above and various network baselines (see Materials and Methods) which have been used in the literature to estimate tie strength.
In Table \ref{tab:hodge_prediction}, we present the cross-validation $R^2$ of models trained on different sets of features.
We find that the Hodge Decomposition features defined above often statistically significantly outperform the network baselines, and are otherwise statistically indistinguishable from the baseline approaches.

\begin{figure}
    \centering
     \hfill
     \begin{subfigure}[b]{0.42\textwidth}
         \centering
         \includegraphics[width=\textwidth]{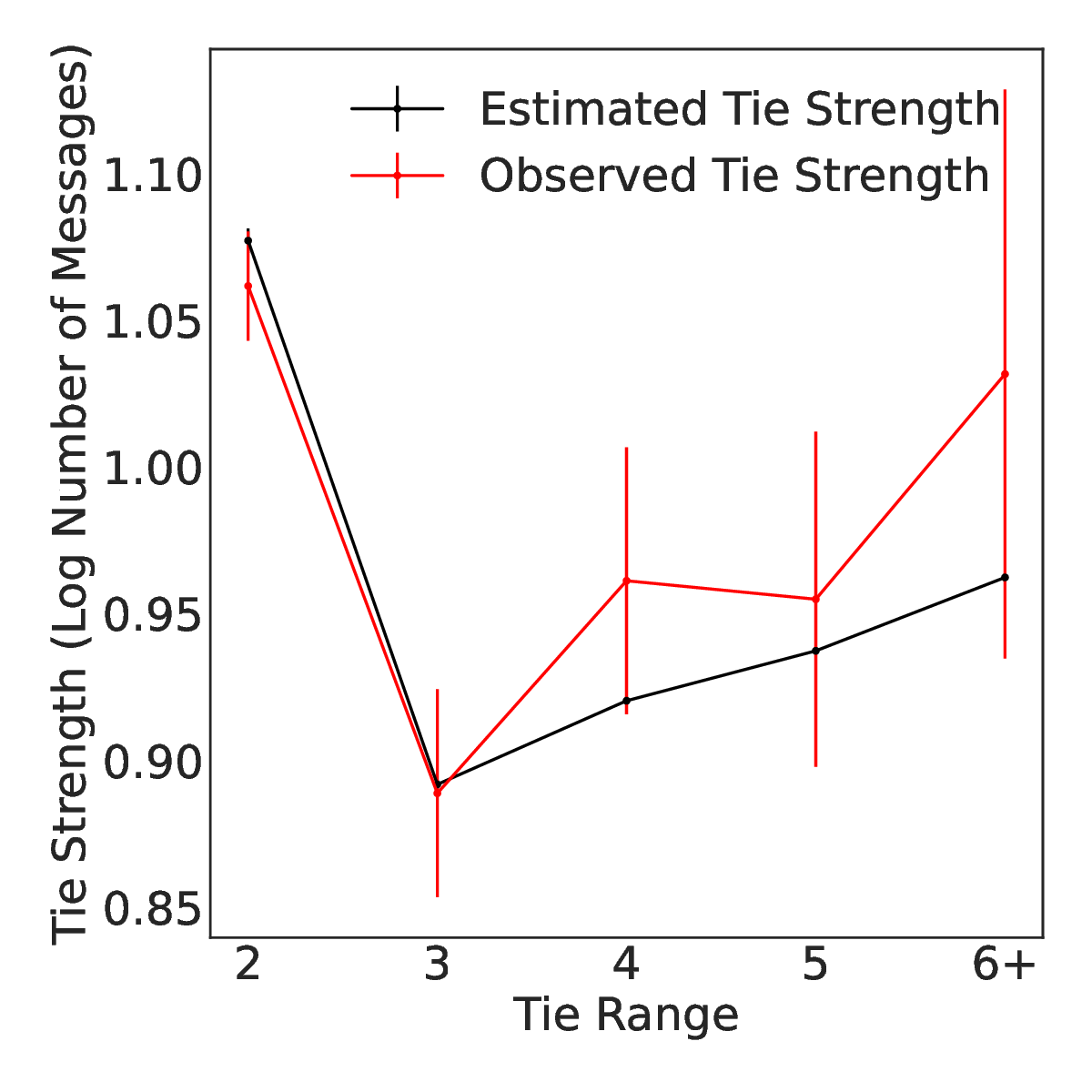}
         \caption{}
         \label{fig:tiestrength_range}
     \end{subfigure}
     \hfill
     \begin{subfigure}[b]{0.42\textwidth}
         \centering
         \includegraphics[width=\textwidth]{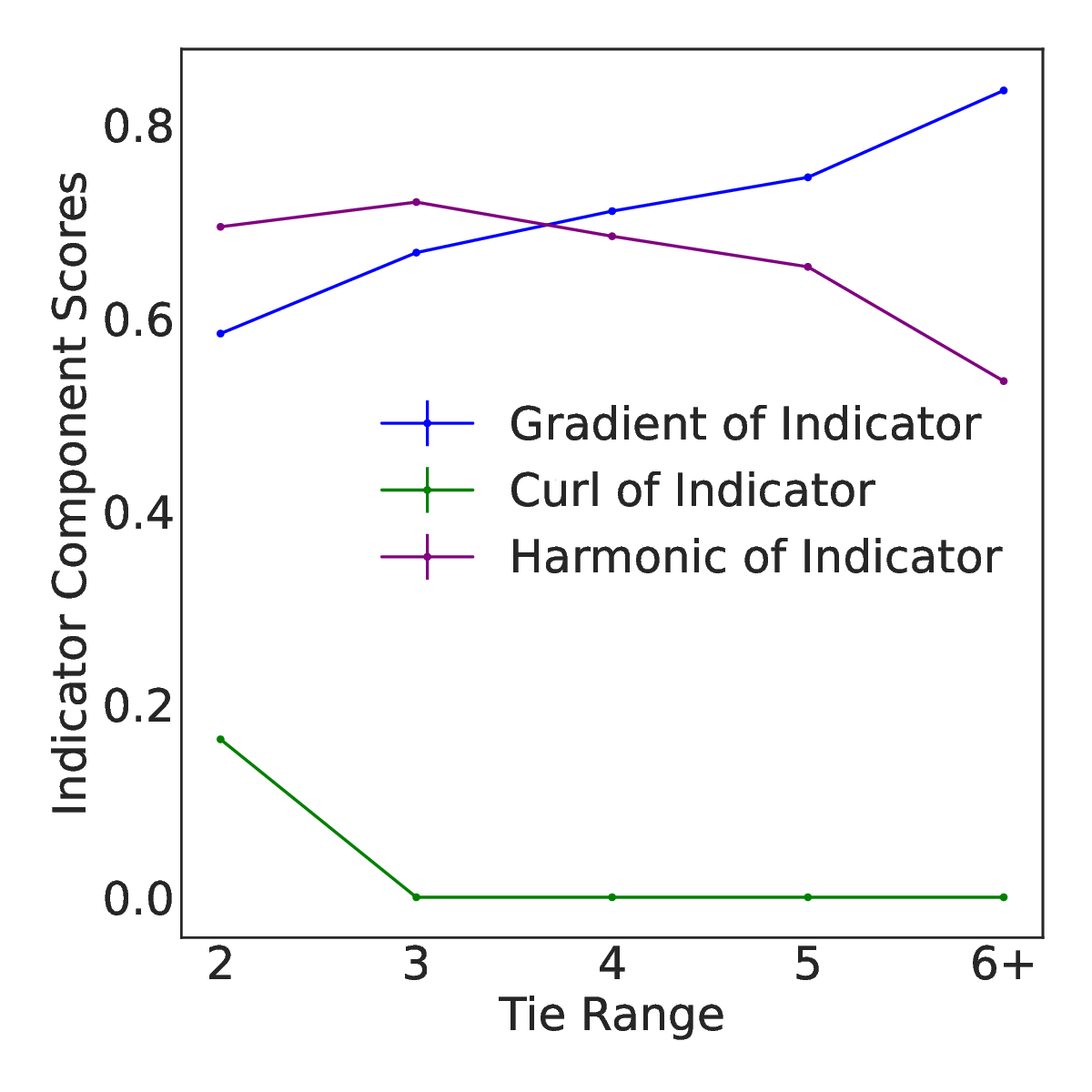}
         \caption{}
         \label{fig:components_range}
     \end{subfigure}
     \hfill
    \caption{Replication of the ``U''-shape with Hodge Decomposition in the \texttt{sms-c} dataset. 
    (a) We find that estimates of tie strength computed using the Hodge Decomposition features replicate the non-monotonic relationship between tie strength and tie range.
    (b) In empirically analyzing the relationship between each Hodge Decomposition feature and tie range, we find that the gradient component increases in tie range, the curl component is only non-trivial for a tie range of 2, and the harmonic component is non-monotonic in tie range. 
    This suggests that tie strength is context dependent. 
    For ties of range 2, the high tie strength can be attributed to the contribution from the curl component, which then drops for all higher tie ranges.
    Then, as tie range increases from 3 onwards, we find that ties with a higher gradient component have a larger weight in the model than that of the harmonic component, resulting in long-range ties being stronger.
    % (a) We find that estimates of tie strength computed using the Edge PageRank measure replicate the strength of long ties. 
    % (b) The Edge PageRank measure itself has an inverted ``U''-shape relationship with tie range, and hence the measure emphasizes weak ties which span medium network distances. Such edges have been theorized to carry non-redundant but useful information \cite{kim2023makes}.
    % (c) The Edge PageRank measure can be seen as a combination of three orthogonal components (gradient, curl, and harmonic) of the indicator vector of an edge (see ``Methods,'' Section \ref{ss:hodge}).
    % In particular, the harmonic component has an inverted ``U''-shape relationship with tie range. 
    % Because the dynamics of Edge PageRank inherently emphasize the harmonic component (Theorem \ref{thm:hodge_informal}), we see that the replication of the ``U''-shape is not simply an artifact of Edge PageRank measuring tie strength quantitatively well. 
    % Rather, we find that Edge PageRank lends itself to a relationship with tie range through its harmonic component.
    }
    \label{fig:u-shape}
\end{figure}

We further find that the model which uses the Hodge Decomposition features is able to replicate the ``U''-shape relationship between tie strength and tie range (Figure \ref{fig:u-shape}). 
As shown in Figure \ref{fig:components_range}, we see that this effect can also be explained by the relationship between each component and tie range. 
Namely, the high tie strength for edges with a tie range of 2 can be attributed to the non-trivial curl component of the indicator, and the high tie strength for edges with higher tie ranges can be attributed to the gradient component.
This suggests that tie strength can be viewed as context dependent.
For edges which share common neighbors, dyadic tie strength is high as expected.
However, for bridging edges, the type of bridge matters--- bridges which are associated with the harmonic subspace (i.e., those which are most related to topological holes) tend to have lower tie strength than those associated with the gradient subspace (i.e., those which are associated with cuts in the graph).

% Figure: Replicating U-shape

\subsection{Theoretical Relationship between Hodge Components and Bridges}

We ultimately find that Hodge components' ability to replicate the ``U''-shape does not simply appear to be an artifact of the measures' quantitative ability to measure tie strength.
That is, one might expect that any set of measures which estimate tie strength well should also capture any stylized fact regarding tie strength.
However, our analysis reveals that each of the gradient, curl, and harmonic components naturally relate to tie range.

We first present an analysis of the gradient component $I_e^g$.
We find that the gradient component is closely related to the notion of a global bridge in a network, where a global bridge is defined as an edge whose removal strictly increases the number of connected components in the network.
Specifically, we have the following claim:
\begin{prop} \label{prop:gradient}
    Consider a simplicial complex $\mathcal{X}$ and an edge $e\in \mathcal{X}$. Then, $I_e^g = 1$ if and only if $e$ is a global bridge.
\end{prop}
The claim is proven in Materials and Methods and follows from the definition of the vector $\delta_e^g$ as being in the image of $B_1^\top$, which conceptually represents the space of cuts in the graph.
Importantly, this indicates that the gradient component is maximized if and only if the edge $e$ is a global bridge, as $I_e^g$ can not exceed 1 by definition.

With respect to tie range, a global bridge can be thought of as a tie whose tie range is infinite, as there is no second shortest path between nodes in a global bridge.
Moreover, because the claim above is an if and only if statement, and because $\delta_e^g$ represents the projection into the cut space, we expect that long-range bridging ties, i.e. those ties with a long second shortest paths, are likely to have high values of $I_e^g$ as well.
Indeed, we are able to empirically validate that ties with higher tie range have larger gradient component values of their indicator vector, on average (see Figure \ref{fig:components_range}).

We next present an analysis of the curl component $I_e^c$. 
We find that an edge can only have a non-trivial curl component if it is incident to at least one filled triangle in the simplicial complex. 
That is, we have the following proposition:
\begin{prop} \label{prop:curl}
    Consider a simplicial complex $\mathcal{X}$ and an edge $e\in \mathcal{X}$. Then, $I_e^c > 0$ if and only if there exists some filled triangle $t \in \mathcal{X}$ such that $e$ is a part of $t$ (equivalently, $e \subset t$.)
\end{prop}
This claim is proven in the Materials and Methods, which notes that $\delta_e^c$ is the projection of $\delta_e$ onto $B_2$, which can be solved for explicitly and involves a term of the form $B_2^\top \delta_e$ which is non-zero if and only if there is a triangle in $\mathcal{X}$ which contains $e$.

The above claim indicates that the curl component of a simplicial complex can only be non-zero if an edge has a tie range of 2, as the edge would have to be a part of a triangle, and we are able to confirm this empirically in Figure \ref{fig:components_range}.

Finally, we consider the harmonic component $I_e^h$. 
We find that this component is most associated with local bridges, which are defined as edges which have a tie range of at least 3.
We formalize this with the following claim:
\begin{prop}\label{prop:harmonic}
    Consider a simplicial complex $\mathcal{X}$ and an edge $e \in \mathcal{X}$. If the tie range of $e$ is finite and at least 3, then $I_e^h > 0$.
\end{prop}
\begin{proof}
The above proposition follows from the previous two, as an edge with finite tie range cannot be a global bridge and hence $I_e^g < 1$ according to Proposition \ref{prop:gradient}, and since the tie range is 3, then $I_e^c = 0$ according to Proposition \ref{prop:curl}. 
Therefore, since $(I_e^g)^2 + (I_e^c)^2 + (I_e^h)^2 = 1$ by definition, it must be the case that $I_e^h > 0$.
\end{proof}
We further note that empirically, the harmonic component $I_e^h$ actually has a non-monotonic relationship with tie range (Figure \ref{fig:components_range}). 
This ultimately makes sense as the harmonic component of an edge is orthogonal to both the curl and gradient components, which are maximized at the two extremes of tie range.

Ultimately, the three theoretical results highlighted above indicate that the proposed structural measures naturally relate to the extent to which a social tie acts as a bridge, and that the replication of the ``U''-shape relationship between tie strength and tie range is not an artifact of the measures' ability to estimate tie strength.

\section{Describing the Strength of Weak Ties with a Single Centrality Measure}

Although the above results replicate and offer an explanation for why certain bridging ties may be strong, they do not resolve the apparent tension between the existence of strong bridging ties and Granovetter's original claim regarding the strength of weak ties.
Towards this end, we provide an analysis of a particular centrality measure, Edge PageRank, which we show is a combination of the three measures discussed above and can also be interpreted through a stochastic information exchange process.
In particular, the measure emphasizes ties that are in a network position to transfer useful information. 

As discussed by Kim and Fernandez \cite{kim2017strength}, two possible ways in which the ``U''-shape relationship between tie range and tie strength can be reconciled with the strength of weak ties thesis are if (1) prior studies of the strength of weak ties thesis were performed on networks too small to observe ties with high tie range and/or (2) medium-distant ties are in the best position to transfer information which is both novel and useful to an individual. 
That is, information from long ties, while novel, may not be as pertinent to an individual.
Our theoretical analysis of the Edge PageRank measure suggests the latter may be true.

% Figure: Explain Edge PageRank
\begin{figure}
    \centering
    \includegraphics[width=\textwidth]{edgepr_ex.eps}
    \caption{Illustration of the Edge PageRank measure. 
    In this figure, we present the ``lifted'' interpretation of the Edge PageRank measure, though the measure can be computed with a direct matrix computation.
    (Lower Left) The example begins with a simplicial complex defined on four nodes. The nodes $\{1, 2, 3\}$ have been a part of a higher-order interaction, whereas nodes $2, 3, $ and $4$ have dyadic relationships but no higher-order interaction. In this example, we will compute the Edge PageRank score for the edge $e = \{2, 3\}$, which we initially represent with the indicator vector $\delta_e$. 
    (Upper Left) The first step in Edge PageRank is to ``lift'' the indicator vector to a vector space which represents each possible direction of each edge, creating the vector $\widehat{\delta_e}$. (Upper Right) Edge PageRank then runs a standard (node) PageRank process in this lifted space. In this PageRank process, the ``teleportation'' vector is taken to be $\widehat{\delta_e}$ and transition probabilities correspond to a graph where directed edges are connected based on their underlying adjacency in the original simplicial complex. The resulting PageRank vector in the lifted space is represented by $\widehat{\pi}_e$.
    (Lower Left) Once the ``lifted'' PageRank process has converged, the resulting values are projected back to the original space of edges by taking the difference between the values corresponding to the two orientations of each edge. This result, $\pi_e$, is referred to as the Edge PageRank vector.
    (Lower Middle) To assign a score to each edge, we take the $2$-norm of the Edge PageRank vector. 
    Although the edges other than $e = \{2, 3\}$ contribute minimally to the Edge PageRank score in this small example, we note that the teleportation parameter and the size of the network both affect the extent to which edges other than $e$ affect its Edge PageRank score.}
    \label{fig:edgepr}
\end{figure}

\subsection{Edge PageRank}

We first define the Edge PageRank measure, and then provide an informal theorem which describes how the measure relates to the Hodge Decomposition features above.
The Edge PageRank measure is defined as a function of a normalized $1$-Hodge Laplacian, defined as follows:

\begin{definition}[Normalized $1$-Hodge Laplacian \cite{schaub2018random}]\label{def:normalized_laplacian}
    For a simplicial complex $\mathcal{X}$ with boundary operators $B_1$ and $B_2$, the normalized $1$-Hodge Laplacian is a matrix $\mathcal{L}_1 \in \mathbb{R}^{|E| \times |E|}$ which satisfies
    \begin{equation}\label{eq:normalized_hodge_laplacian}
        \mathcal{L}_1 = D_2 B_1^\top D_1^{-1} B_1 + B_2 D_3 B_2^\top D_2^{-1} \,,
    \end{equation}
    where $D_1$, $D_2$, and $D_3$ are diagonal normalization matrices such that $D_3[t, t] = 1/3$ for all triangles, $D_2[e, e] = \max \{\deg(e), 1\}$ for all edges, where the degree of an edge is the number of filled triangles associated with that edge, and $D_1[v, v] = 2 \sum_{e|v \in e} D_2[e, e]$ is a weighted node degree. 
\end{definition}

As discussed by Schaub \textit{et al.} \cite{schaub2018random}, this version of the $1$-Hodge Laplacian can be interpreted as a random walk operator in a ``lifted'' space of edges, which they formalize in the following Theorem.
\begin{theorem}[Stochastic Lifting of $\mathcal{L}_1$ \cite{schaub2018random}] \label{thm:hodge_lifting}
    Consider a simplicial complex $\mathcal{X}$ with $|E|$ edges, and define the lifting operator $V \in \mathbb{R}^{2|E| \times |E|}$ as
    \begin{equation} \label{eq:lifting}
        V = \begin{bmatrix}
        ~~I_{|E|} \\ -I_{|E|}
        \end{bmatrix} \,.
    \end{equation}
    The normalized $1$-Hodge Laplacian $\mathcal{L}_1$ satisfies
    \begin{equation} \label{eq:hodge_laplacian_lifting}
        -\frac{1}{2} \mathcal{L}_1 V^\top = V^\top \hat{P} \,,
    \end{equation}
    where $\hat{P}\in \mathbb{R}^{2|E| \times 2|E|}$ is a stochastic matrix whose sparsity depends on the structure of the simplicial complex.
\end{theorem}
% The authors show how this random walk can be augmented to describe an algebraic topological extension of the classical PageRank measure, which they refer to as Edge PageRank.
The definition of the normalized $1$-Hodge Laplacian ultimately allows us to define the Edge PageRank measure for a particular edge:
\begin{definition}[Edge PageRank \cite{schaub2018random}] \label{def:edgepr}
    Consider a simplicial complex $\mathcal{X}$ with normalized $1$-Hodge Laplacian $\mathcal{L}_1$, an edge $e \in E$, and a constant $\beta \in (2, \infty)$.
    Denote the indicator vector corresponding to edge $e$ as $\delta_e \in \mathbb{R}^{|E|}$, which has a $1$ at the entry corresponding to edge $e$ and is equal to $0$ elsewhere.
    The Edge PageRank score $EP_e$ is defined:
    \begin{equation} \label{eq:edgepr_norm}
        EP_e = \| \pi_e \| \,,
    \end{equation}
    where $\pi_e$ satisfies
    \begin{equation} \label{eq:personalized_epr}
        (\beta I +  \mathcal{L}_1) \pi_e = (\beta - 2) \delta_e \,.
    \end{equation}
\end{definition}

The Edge PageRank score has an equivalent description as a stochastic process in the lifted space of edges, where each edge $\{i, j\}$ is represented by two directed edges $[i, j]$ and $[j, i]$, as described in Figure \ref{fig:edgepr} and noted in the following Proposition proven by Schaub \textit{et al.} \cite{schaub2018random}.

\begin{prop} \label{prop:edgepr_lifting}
    For a simplicial complex $\mathcal{X}$ with normalized 1-Hodge Laplacian $\mathcal{L}_1$, an edge $e \in E$, and a constant $\beta \in (2, \infty)$, let the Edge PageRank vector $\pi_e$ be defined as in \eqref{eq:personalized_epr}. The Edge PageRank vector satisfies the following equation:
    \begin{equation} \label{eq:edgepr_lifted}
        \pi_e = V^\top \hat{\pi}_e \,,
    \end{equation}
    where $V$ is as in \eqref{eq:lifting} and $\hat{\pi}_e$ is a probability vector which satisfies 
    \begin{equation}
        \left(I_{2|E|} - \alpha \hat{P}\right) \hat{\pi}_e = (1 - \alpha) \hat{\delta}_e \,,
    \end{equation}
    for $\alpha = 2/\beta$, $\hat{\delta}_e$ as the indicator vector for the edge $e$ which is $1$ for a particular orientation of $e$ and $0$ elsewhere, and $\hat{P}$ is that of Theorem \ref{thm:hodge_lifting}
\end{prop}

The above proposition suggests that $\pi_e$ can be interpreted as the projected version of a standard PageRank vector $\hat{\pi}_e$.
Further, the result indicates that the parameter $\beta$ in Definition \ref{def:edgepr} corresponds to the teleportation parameter in the PageRank process defined in the lifted space.
In experiments, $\beta = 2.5$ is used.
The above proposition also reveals that the Edge PageRank score defined above is effectively a ``personalized'' Edge PageRank score, since the indicator vector corresponding to an edge is used to define the measure.
This is done for technical reasons, and ultimately defining the Edge PageRank score in this way avoids any possibility that the ordering of the nodes (i.e., the labeling of the nodes with natural numbers) affects the score (see \cite{schaub2018random}, Section 6.2).

\begin{figure}
    \centering
    \includegraphics[width=0.3\textwidth]{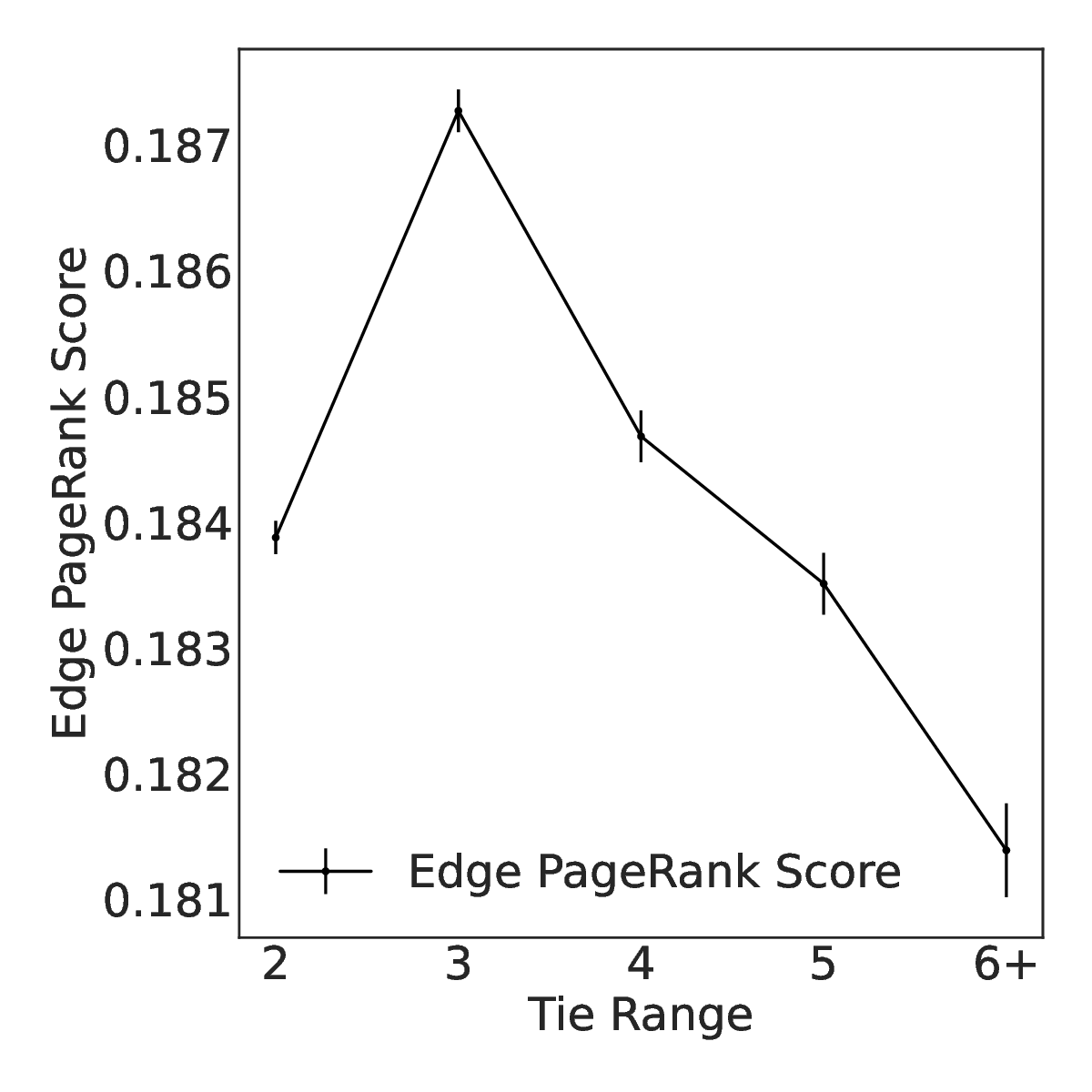}
    \caption{Edge PageRank scores as a function of tie range in the \texttt{sms-c} dataset. We find that the Edge PageRank measure, in emphasizing the harmonic component of the indicator, also has a non-monotonic relationship with tie range and hence can ultimately capture the ``U''-shape relationship between tie strength and tie range. }
    \label{fig:edgepr_range}
\end{figure}

Our theoretical analysis of this measure reveals that it can be written as a function of a weighted version of the Hodge Decomposition mentioned in the previous section.
In particular, we present the following (informal) Theorem.
\begin{theorem}\label{thm:hodge_informal}
    (Informal) Consider an edge $e$ in a simplicial complex $\mathcal{X}$. 
    The Edge PageRank vector $\pi_e$ can be written as a function of a weighted Hodge Decomposition of the indicator vector $\delta_e$, such that the Edge PageRank vector emphasizes the harmonic component of $\delta_e$ as opposed to the gradient and curl components of $\delta_e$.
\end{theorem}
The Theorem is formally stated and proven in Materials and Methods.
Because Edge PageRank emphasizes the harmonic component of the indicator of an edge, we find that Edge PageRank itself has an inverted ``U''-shape relationship to tie range (Figure \ref{fig:edgepr_range}).
Further, since the Edge PageRank function is inversely related to tie strength, we see that this allows the measure to replicate the ``U''-shape relationship between tie strength and tie range.

Moreover, we note that the Edge PageRank measure is also an effective measure of tie strength in its own right, as noted in Table \ref{tab:pagerank_tiestrength}.
In particular, the measure outperforms or is statistically indistinguishable from other proxies of tie strength in 11 out of 15 datasets.
We find that Edge PageRank is highly inversely correlated with tie strength.
Across all edges in all datasets, the Edge PageRank measure and tie strength have a correlation coefficient of $-0.291$ ($p < 10^{-16}$).
This ultimately suggests that there is a sense in which Edge PageRank, as a measure of centrality, emphasizes the structural properties of weak ties which make them important, which we interpret in the following section.

% Table: Tie Strength Prediction Results
\begin{table}[]
    \centering
    \caption{Accuracy of predicting tie strength with a single regressor. Each entry is the test accuracy of a linear regression, computed using a 10-fold cross-validation. 
    Bolded entries indicate when a measure statistically significantly outperforms all others.}
    \footnotesize
    \begin{tabular}{rrrrr}
\toprule
&                             Edge PageRank &                        \shortstack[r]{Unweighted\\Overlap} &                                Dispersion &                       Betweenness \\
\midrule
\texttt{bills-house}            &           0.044 \scriptsize{($\pm$0.001)} &  \textbf{0.098} \scriptsize{($\pm$0.004)} &           0.055 \scriptsize{($\pm$0.003)} &   0.001 \scriptsize{($\pm$0.000)} \\
\texttt{bills-senate}           &           0.120 \scriptsize{($\pm$0.004)} &  \textbf{0.157} \scriptsize{($\pm$0.005)} &           0.004 \scriptsize{($\pm$0.002)} &   0.011 \scriptsize{($\pm$0.001)} \\
\texttt{coauth-dblp}            &           0.008 \scriptsize{($\pm$0.000)} &           0.007 \scriptsize{($\pm$0.000)} &  \textbf{0.310} \scriptsize{($\pm$0.002)} &   0.005 \scriptsize{($\pm$0.000)} \\
\texttt{college-msg}            &           0.008 \scriptsize{($\pm$0.001)} &           0.018 \scriptsize{($\pm$0.002)} &           0.006 \scriptsize{($\pm$0.001)} &   0.015 \scriptsize{($\pm$0.002)} \\
\texttt{contact-high-school}    &  \textbf{0.390} \scriptsize{($\pm$0.011)} &           0.187 \scriptsize{($\pm$0.007)} &          -0.002 \scriptsize{($\pm$0.001)} &   0.097 \scriptsize{($\pm$0.007)} \\
\texttt{contact-hospital}       &  \textbf{0.263} \scriptsize{($\pm$0.021)} &           0.182 \scriptsize{($\pm$0.019)} &          -0.062 \scriptsize{($\pm$0.056)} &   0.067 \scriptsize{($\pm$0.011)} \\
\texttt{contact-primary-school} &  \textbf{0.522} \scriptsize{($\pm$0.004)} &           0.309 \scriptsize{($\pm$0.006)} &          -0.002 \scriptsize{($\pm$0.000)} &   0.138 \scriptsize{($\pm$0.006)} \\
\texttt{contact-university}     &  \textbf{0.203} \scriptsize{($\pm$0.002)} &           0.116 \scriptsize{($\pm$0.002)} &          -0.000 \scriptsize{($\pm$0.000)} &   0.002 \scriptsize{($\pm$0.000)} \\
\texttt{contact-workplace-13}   &  \textbf{0.226} \scriptsize{($\pm$0.008)} &           0.089 \scriptsize{($\pm$0.008)} &          -0.006 \scriptsize{($\pm$0.003)} &   0.025 \scriptsize{($\pm$0.004)} \\
\texttt{contact-workplace-15}   &  \textbf{0.218} \scriptsize{($\pm$0.004)} &           0.133 \scriptsize{($\pm$0.004)} &          -0.002 \scriptsize{($\pm$0.001)} &  -0.103 \scriptsize{($\pm$0.116)} \\
\texttt{email-Enron}            &  \textbf{0.288} \scriptsize{($\pm$0.023)} &           0.166 \scriptsize{($\pm$0.012)} &          -0.004 \scriptsize{($\pm$0.008)} &  -0.008 \scriptsize{($\pm$0.006)} \\
\texttt{email-Eu}               &  \textbf{0.192} \scriptsize{($\pm$0.003)} &           0.146 \scriptsize{($\pm$0.006)} &           0.011 \scriptsize{($\pm$0.002)} &   0.017 \scriptsize{($\pm$0.002)} \\
\texttt{india-villages}         &  \textbf{0.674} \scriptsize{($\pm$0.001)} &           0.482 \scriptsize{($\pm$0.001)} &           0.001 \scriptsize{($\pm$0.000)} &   0.182 \scriptsize{($\pm$0.002)} \\
\texttt{sms-a}                  &           0.000 \scriptsize{($\pm$0.000)} &           0.011 \scriptsize{($\pm$0.001)} &           0.009 \scriptsize{($\pm$0.001)} &  -0.000 \scriptsize{($\pm$0.000)} \\
\texttt{sms-c}                  &          -0.001 \scriptsize{($\pm$0.000)} &  \textbf{0.014} \scriptsize{($\pm$0.003)} &           0.002 \scriptsize{($\pm$0.001)} &  -0.001 \scriptsize{($\pm$0.000)} \\
\bottomrule
\end{tabular}
    \label{tab:pagerank_tiestrength}
\end{table}

\subsection{Interpreting the Underlying Stochastic Process}
Many measures of centrality on a social network can be understood through an underlying social process~\cite{bonacich1987power,friedkin1990social,friedkin1991theoretical}. 
For instance, when Katz \cite{katz1953new} introduced a centrality measure on networks as a mathematical function of the adjacency matrix of the network, the measure required a hyperparameter for convergence and this hyperparameter was not given any social significance.
Bonacich \cite{bonacich1987power} later provided a social interpretation of this hyperparameter by contextualizing the measure introduced by Katz with a social process.
Similarly, Friedkin and Johnsen \cite{friedkin1990social} showed that several centrality measures which had previously been described as equilibria could also be interpreted through a social process.

The present section analyzes the Edge PageRank measure under this lens of analysis. 
We show that the Edge PageRank measure, which had previously been defined as a purely mathematical extension of the classical PageRank measure, can be interpreted through an underlying social process.
Specifically, we show how the vector $\hat{\pi}_e$ in Eq. \eqref{eq:edgepr_lifted} can be interpreted as the equilibrium of a stochastic communication process.

\begin{figure}
    \centering
    \includegraphics[width=0.6\textwidth]{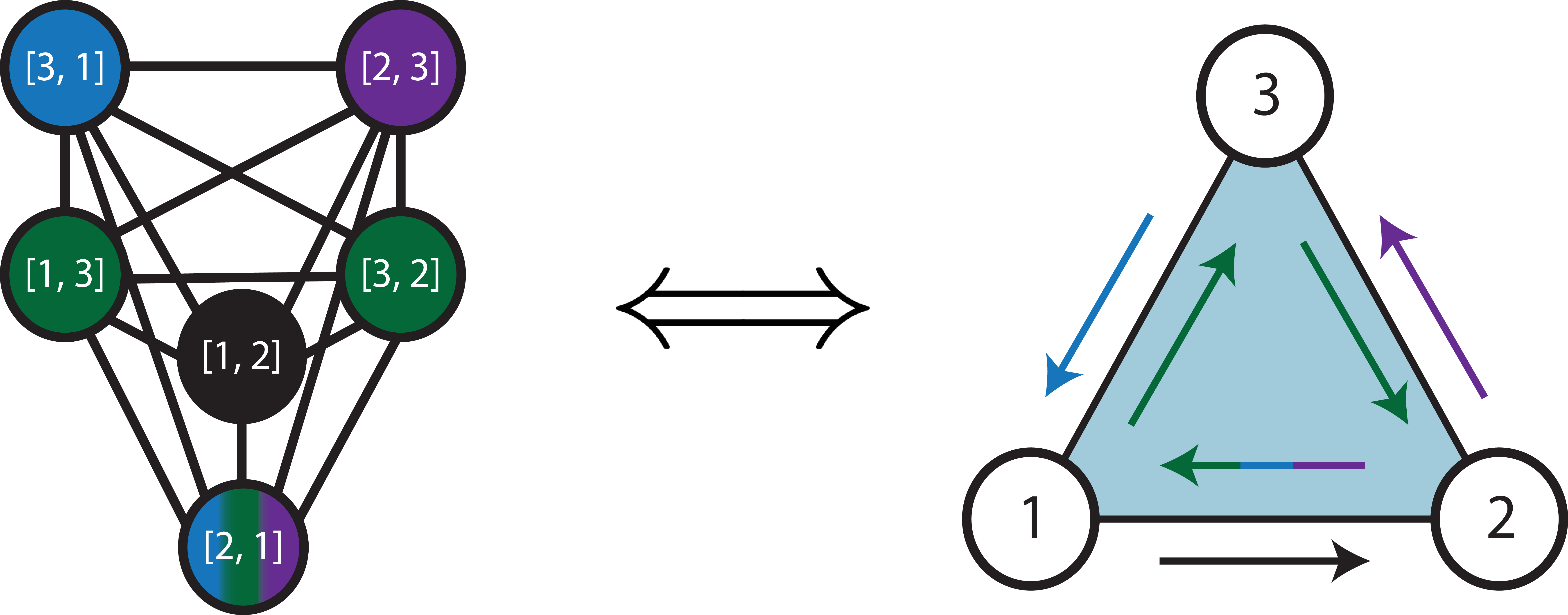}
    \caption{Illustration of the interpretation of the Edge PageRank diffusion as a communication process. Each state of the process corresponds to a directed edge, which we interpret as passed messages. If the state of the random walk is currently $[1, 2]$ (black), i.e. node $1$ has just sent a message to node $2$, then there are three possible types of steps that can be taken next. 
    The lower walk (purple) indicates that, after node $2$ receives a message, then node $2$ may send a message to one of its neighbors, including node $1$.
    This walk is designed to be reversible (blue), due to the lifted graph being undirected, which we interpret as node $1$ seeking information after sending a message to node $2$.
    The upper walk (green) represents the effect of higher order information in the process, and enables two communication possibilities which are natural in a co-present interaction. 
    Once node $1$ sends a message to node $2$, then $1$ can send a message to node $3$, as if a message is being sent to both members, or node $3$ might send a message to $2$, as a reaction to node $1$'s message to node $2$.
    We note that if the triangle were unfilled, then neither green arrow would be possible.}
    \label{fig:lifted_ex}
\end{figure}

\paragraph{The Stochastic Communication Process} 

To interpret the lifted PageRank procedure used to define $\hat{\pi}_e$, we first provide an interpretation of the states of the lifted graph. 
We then discuss how the random process will ``walk'' (transition or move) between these states, and $\hat{\pi}_e$ will represent the likelihood of each state occurring at equilibrium.

The lifted graph contains two states (nodes) for every edge $\{i, j\}$ in the original simplicial complex--- a state $[i, j]$ and a state $[j, i]$. 
We will interpret the stochastic process being at state $[i, j]$ as the node $i$ sending a message to node $j$.
Hence, as the diffusion process transitions between different states, we can interpret the sequence of states as a sequence of messages being sent from one node to another.

The diffusion process is defined with respect to a seed edge $e = \{s_i, s_j \}$, where without loss of generality we assume $s_i < s_j$, and the initial state of the diffusion is where the state is equal to $[s_i, s_j]$, i.e. we assume that the initial state of the stochastic process is that node $s_i$ sends a message to node $s_j$.
With probability $1 - \alpha$, where $\alpha = 2/\beta$ in the definition of Edge PageRank (see Proposition \ref{prop:edgepr_lifting}), the random walk will teleport back to the seed edge and $s_i$ sends another message to $s_j$.
With probability $\alpha$, the stochastic process will transition according to the matrix $\hat{P}$ defined in Theorem \ref{thm:hodge_lifting}.
The matrix $\hat{P}$ takes the following form:
\[ \hat{P} = \frac{1}{2}\hat{P}_{lower} + \frac{1}{2}\hat{P}_{upper}\,,\]
such that, conditional on the stochastic process transitioning according to $\hat{P}$, with probability $1/2$ the next message will be passed according to the lower walk and with probability $1/2$ the next message will be sent according to the upper walk. 
After the stochastic process transitions, the procedure repeats from this new state, and the process either returns to the seed state with probability $1 - \alpha$ or takes another step according to $\hat{P}$ with probability $\alpha$.

The lower walk represents the standard type of message passing often studied in graph-based networks--- if the current state is $[i, j]$, i.e. $i$ has just sent a message to $j$, then the next possible steps of the lower walk are of the form $[j, k]$ where $k$ is a neighbor of $j$ in the usual sense (it is possible $k = i$).
That is, after $i$ sends a message to $j$, $j$ will send a message to one of its neighbors.
As discussed in Schaub \textit{et al.} \cite{schaub2018random}, the selection of the neighbor $k$ is taken proportionally to the upper degree of the edge of $\{j, k\}$, i.e. $j$ is more likely to send the message to a neighbor who is a part of many filled triangles with $j$.
Because the lifted graph is undirected, the lower walk is defined to be reversible, which here we interpret as a node seeking information after sending a message. 
It's worth noting that a potential extension of the present method would be to consider how the measure may change if we do not allow for the lower walk to be reversible, as the stochastic process may become more interpretable.
However, we leave the lower walk reversible in this work as it allows for the theoretical decomposition of the Edge PageRank measure in Theorem \ref{thm:hodge_informal}.

The upper walk is the aspect of the Edge PageRank process which encodes higher-order information.
If nodes $i$, $j$, and $k$ are a part of a filled triangle, then there are additional possible transitions possible after the message $[i, j]$ is passed. 
In particular, the presence of the triad allows the walk to transition to $[i, k]$ or $[k, j]$, both of which are reasonable to expect in a group conversation.
For example, if $i$ sends a message to $j$ and $k$ is also present, it is reasonable to expect that $i$ may also be sending a message to $k$.
Alternatively, if $i$ sends a message to $j$, then $k$ may also choose to send a message to $j$ as well, having seen $i$ send a message to $j$.
In this sense, the co-presence of $i, j,$ and $k$ leads to new communication pathways, which is natural to expect in social settings (c.f. \cite{bianchi2024relational}).

\subsubsection{Discounting Filled Triangles and Emphasizing Short Cycles} 
The stochastic process above reveals how Edge PageRank discounts filled triangles and emphasizes cycles, causing it to highlight medium distance ties who provide non-redundant and useful information, respectively.
Mathematically, these ideas are formalized in Theorem \ref{thm:hodge_informal}, which indicates that Edge PageRank discounts filled triangles by de-emphasizing the curl component and emphasizes cycles by emphasizing the harmonic component relative to the gradient component.

The description of the upper walk shows how the existence of higher-order interactions causes $\hat{\pi}_e$ to be more uniform across different orientations of each edge (see Figure \ref{fig:lifted_ex}), and this ultimately leads to $V^\top \hat{\pi}_e = \pi_e$ having a lower magnitude.
Hence, the stochastic process underlying Edge PageRank ultimately provides less weight to edges which take part in filled triangles.
Moreover, the description of the lower walk shows that Edge PageRank emphasizes cycles which do not have filled triangles--- in particular, if there exists a cycle in the simplicial complex, then it is possible that the lower walk can traverse the entire cycle in one orientation, creating an imbalance which is emphasized by Edge PageRank. 
The probability of the lower walk traversing an entire cycle is highest for short cycles, which also correspond to topological holes in the network when these cycles are absent of filled triangles.
That is, our interpretation of the information exchange process further reveals how the Edge PageRank measure emphasizes the harmonic component as opposed to the curl or gradient components of an edge, thereby identifying topological obstructions or bottlenecks which may be conduits for useful information.

\section{Discussion} 
\label{s:discussion}

We have introduced structural measures based on the theory of Hodge Decomposition which outperform standard network-based proxies in estimating dyadic tie strength in social networks. 
These measures further capture a sociological puzzle noted in the literature regarding how long ties can be surprisingly strong, and offer an explanation in that dyadic tie strength can be context dependent.

Importantly, our results reconcile the apparent dissonance between the result that long-range ties can be strong and Granovetter's original thesis about the strength of weak ties.
As argued by Kim and Fernandez \cite{kim2023makes},  the idea of long, bridging ties which are strong need not be incommensurate with the idea that weak ties can provide valuable information due to their network position.
In particular, they suggest that it could be the case that either most datasets on tie strength are right-truncated, i.e. most prior datasets were too small to say anything substantive about long-range ties, or that it is possible that medium-distance weak ties are likely to provide valuable information.

Our analysis of a single centrality measure, Edge PageRank, favors the latter proposed reconciliation.
Our findings connect the Edge PageRank measure of Schaub \textit{et al.}  \cite{schaub2018random} to both our proposed network measures and a model of information exchange which highlights ties which are posed to transfer useful information.
Theoretically, we find that the Edge PageRank measure discounts both filled triangles, for which additional communication can lead to information redundancies, and extremely long ties, across which information may be novel but not as relevant to an individual.
We further validate that the Edge PageRank measure estimates tie strength well in empirical networks and captures a non-monotonic relationship between tie range and tie strength.

These results carry consequences for studies of social networks more broadly. 
Our approach explicitly encodes higher-order information about interactions where three or more individuals are co-present, which we find is extremely valuable for estimating tie strength.
This suggests that data on social structure should be collected keeping such co-present settings in mind, as opposed to collecting data at the level of dyadic relationships.
Moreover, our use of tools from algebraic topology to understand global social structure suggests that these tools may be of value to sociologists.
For example, the use of Hodge Decomposition to analyze the network context of an edge captures notions of topological holes, which may be related to theories of structural holes and social capital in the sociology literature more broadly \cite{burt2002social}.

Tie strength has remained an influential construct in the social sciences for over half of a century, as it provides an intuitive way to relate micro-level interactions to macro-level outcomes \cite{granovetter1973strength,mattie2018understanding,kim2023makes}.
By augmenting tie strength prediction with information from higher-order interactions, we have provided initial evidence that incorporating higher-order information can be valuable for understanding the nature of micro-level interactions.
Potential future research can build on the present study to validate the claims about the proposed network measures and the Edge PageRank measure by analyzing the specific content and outcomes associated with different edges in a network.
By building on recent literature on information novelty \cite{aral2022what}, economic outcomes \cite{jahani2023long,rajkumar2022causal}, and dynamic models of higher-order interactions \cite{iacopini2019simplicial}, we may better understand the mechanisms by which higher-order interactions shape macro-level outcomes such as job placement, socioeconomic status, and economic resilience.

\bibliographystyle{Science}

\bibliography{scibib}

% \section{Acknowledgments}
% The authors thank Ben Golub, Matthew Jackson, Patrick Park, and Amin Rahimian for helpful conversations and comments about this work. 
% This work was supported in part by funding  from  a Vannevar Bush Fellowship from the Office of the Secretary of Defense and from Army Research Office Multidisciplinary University Research Initiative W911-NF-19-1-0217.

\section{Supplementary materials}
Materials and Methods\\

% For your review copy (i.e., the file you initially send in for
% evaluation), you can use the {figure} environment and the
% \includegraphics command to stream your figures into the text, placing
% all figures at the end.  For the final, revised manuscript for
% acceptance and production, however, PostScript or other graphics
% should not be streamed into your compliled file.  Instead, set
% captions as simple paragraphs (with a \noindent tag), setting them
% off from the rest of the text with a \clearpage as shown  below, and
% submit figures as separate files according to the Art Department's
% instructions.

\clearpage

\appendix

\section{Materials and Methods}

\subsection{Definitions of Network Baselines}

In Table  \ref{tab:hodge_prediction} and Table \ref{tab:pagerank_tiestrength} of the main text, we compare algebraic topological network measures to network-based baselines computed using the underlying graph of the data.

As proxies for tie strength in Table \ref{tab:pagerank_tiestrength}, we consider the unweighted overlap measure defined by Onnela \textit{et al.} \cite{onnela2007structure}, the dispersion measure defined by Backstrom and Kleinberg \cite{backstrom2014romantic}, and the betweenness measure formally defined by Freeman \cite{freeman1977set}.
Each of these measures is a function of the graph $G = (V, E)$ for each dataset, where $V$ is the set of nodes and $E$ represents the set of pairwise relationships observed in the data.

\subsubsection{Individual Proxies of Tie Strength}
\paragraph{Unweighted Overlap}
The unweighted overlap measure was used by Onnela \textit{et al.} as a proxy for tie strength, and is defined as follows:

\begin{definition}[Unweighted Overlap]
    For a graph $G = (V, E)$ and an edge $e = \{i, j\} \in E$, the unweighted overlap of the edge $e$ is defined:
    \[ O_{e} = \frac{n_{ij}}{k_i + k_j - n_{ij} - 2} \,,\]
    where $k_i$ and $k_j$ are the degree of nodes $i$ and $j$, respectively, and $n_{ij}$ represents the number of common neighbors between nodes $i$ and $j$.
\end{definition}

The measure has the property that if nodes $i$ and $j$ have no mutual friends, then $O_{e} = 0$, and if all friends are commonly shared, then $O_{e} = 1$.
The intuition that the unweighted overlap measure should approximate tie strength follows that of Granovetter \cite{granovetter1973strength}, who suggests that strong ties should be associated with overlapping social circles.

\paragraph{Dispersion}
The dispersion measure has been used by Backstrom and Kleinberg to identify a particular type of strong ties due to romantic relationships \cite{backstrom2014romantic}.
The measure is defined as follows:
\begin{definition}[Dispersion]
    For a graph $G = (V, E)$ and an edge $e = \{i, j\} \in E$, let
    \[ \textrm{disp}_i(j) = \sum_{s, t \in C_{i}(j)} d_j(s, t) \,,\]
    where $C_{i}(j)$ represents all pairs of nodes which are neighbors of node $i$ other than $i$ and $j$ themselves, and $d_j$ is a distance function described below.
    The dispersion of an edge is then defined:
    \[ D_{e} = \frac{\textrm{disp}_i(j) + \textrm{disp}_j(i)}{2 n_{ij}} \,,\]
    where $n_{ij}$ is again the number of common neighbors between nodes $i$ and $j$. 
\end{definition}
Many choices of distance function can be used to define dispersion, and here we use the same distance function defined by Backstrom and Kleinberg \cite{backstrom2014romantic}--- $d_j(s, t)$ is equal to $1$ if $s$ and $t$ do not share an edge and share no common neighbors with $i$ other than $i$ and $j$ themselves, and $0$ otherwise.

The use of dispersion as an estimate of tie strength is based on the notion of \emph{social foci}, which states that an individual's social connections can often be categorized into clusters.
Particularly strong ties, such as those due to marriage, would be more likely to have been introduced to friends from multiple foci.

In Backstrom and Kleinberg \cite{backstrom2014romantic}, dispersion is an asymmetric measure, as the distance measure is defined relative to each ego. 
Here, we consider a symmetric version of the measure.
We also normalize by the number of common neighbors between $i$ and $j$, as is done in the original reference, with the convention that $D_{e} = 0$ if $i$ and $j$ share no common neighbors.

\paragraph{Betweenness}
The final individual network-based baseline we consider is betweenness, as formalized by Freeman \cite{freeman1977set} and used by Krackhardt \textit{et al.} \cite{krackhardt2003strength} as a network measure expected to be inversely correlated with tie strength.

The betweenness centrality measure for an edge is defined as follows:
\begin{definition}[Betweenness Centrality]
For a graph $G = (V, E)$ and an edge $e = \{i, j\} \in E$, betweenness centrality is defined
    \[ B_{e} = \sum_{s \in V} \sum_{t \in V \setminus \{s\}} \frac{\sigma_{st}(e)}{\sigma_{st}} \,,\]
where $\sigma_{st}$ is the number of shortest paths between nodes $s$ and $t$, and $\sigma_{st}(e)$ is the number of shortest paths between $s$ and $t$ which specifically go through the edge $e$.
\end{definition}
It is worth noting that this particular definition of betweenness centrality is that defined for edges, i.e. it is the \emph{edge betweenness centrality} as opposed to the node betweenness centrality.
As discussed by Krackhardt \textit{et al.} \cite{krackhardt2003strength}, this centrality measure would be expected to be inversely related to tie strength, as it relates to Granovetter's notion of a local bridge--- namely, edges with high betweenness are likely to be connecting individuals in disparate parts of the network, much like local bridges.
Hence, because Granovetter argued that local bridges are likely to be weak ties, Krackhardt \textit{et al.} suggest an inverse correlation between tie strength and betweenness centrality.

\subsubsection{Baselines with Multiple Network Features}

\paragraph{Network Baseline}
In Table \ref{tab:hodge_prediction}, we consider a network baseline which uses three measures to predict tie strength which have been shown to be the best three for predicting tie strength in the context of the same \texttt{india-villages} dataset used in this work as well as a large-scale phone call network \cite{mattie2018understanding}. 
These three measure are the sum of degrees of the nodes on each edge, the unweighted overlap measure defined above, and the sum of clustering coefficients of each node on an edge.
Here, the clustering coefficient for a node is defined as follows:
\begin{definition}[Clustering Coefficient]
For a graph $G = (V, E)$ and a node $i \in V$, the clustering coefficient is defined
     \[ C_i = \frac{2 \big\vert \{j, k\}  ~|~ \{i, j\} \in E, \{i, k \} \in E, \{ j, k \} \in E \big\vert }{k_i (k_i - 1)} \,.\]
Equivalently, this is the fraction of pairs of neighbors that $i$ has who share an edge.
\end{definition}

\subsubsection{Node PageRank Baseline}

We also consider a suite of measures which use the traditional node PageRank measure to estimate tie strength, where PageRank is defined as in \cite{gleich2015pagerank}.
In this baseline, we consider four network measures based on transforming the standard node PageRank scores into measures defined on edges.

The first considered measure is to assign each edge the average PageRank score of the two incident nodes, where PageRank is computed using the standard uniform teleportation vector.
To create the second measure, we compute the personalized node PageRank vector of each node and summarize with the $2$-norm, to consider a measure which is similar to edge PageRank.
For each edge, the second considered measure is then the average of the $2$-norm of personalized PageRank for the two incident nodes.

The third considered measure is the node PageRank score of the line graph. 
Specifically, we convert the graph $G = (V, E)$ into a new graph $L$ where each edge in $E$ becomes associated with a node in $L$, and two nodes in $L$ are connected if their associated edges in $G$ shared a node.
We compute the standard PageRank score on $L$ using the uniform teleportation vector, and then the third considered measure for each edge is that edge's associated PageRank score in the line graph.
Finally, the fourth measure is the $2$-norm of the personalized PageRank vector for each edge's corresponding node in the line graph.

\subsection{Experimental Details}
For the experiments on tie strength, we denote the dependent variable as $y_e$ for each edge, which is a real number which measures the tie strength of the edge. 
We then compute a set of independent variable for each edge based on the network topology, either using the Hodge Decomposition measures in Definition \ref{def:hodge_measures}, the Edge PageRank score in Eq. \eqref{eq:edgepr_norm} or baselines from the literature described in the Supplementary Material.

We use validation accuracies and their standard deviations in a 10-fold cross-validation to report accuracies in Table \ref{tab:hodge_prediction} and Table \ref{tab:pagerank_tiestrength}.
Specifically, each independent variable is used to learn a linear model from the training set, which allows for estimates of tie strength $\hat{y}_e$ to be computed for each edge.
Then, on the held out of edges $E_{held}$, we compute a validation accuracy $ 1 - \frac{1}{|E_{held}|}\sum_{e \in E_{held}} (\hat{y}_e - y_e)^2$ to measure the out-of-sample accuracy of the prediction.

\subsection{Theoretical Results on Hodge Decomposition and Tie Range}

\paragraph{Proof of Proposition \ref{prop:gradient}}

Given a simplicial complex $\mathcal{X}$ and an edge $e$ which belongs to $\mathcal{X}$, we wish to prove that $I_e^g = 1$ if and only if $e$ is a global bridge. 
Without loss of generality, $e$ can be written as $e = \{ i, j \}$ for $i < j$, where $i$ and $j$ are the two nodes associated with the edge (recall that we assign an arbitrary numbering of nodes from $1$ to $|V|$).

$(\Rightarrow)$ Suppose $I_e^g = 1$. Then, it must be the case that $\delta_e^g = \delta_e$. This follows from the fact that $\| \delta_e^g\|^2 + \| \delta_e^c\|^2 + \|\delta_e^h\|^2 = 1$ and the definition $I_e^g = \| \delta_e^g\|$, which ultimately implies that $\delta_e^c = \delta_e^h = 0$.

Since $\delta_e^g = \delta_e$, it must be the case that $\delta_e \in \textrm{im} (B_1^\top)$. 
Hence, there exists some node-dimensional vector $v \in \mathbb{R}^{|V|}$ such that $B_1^\top v = \delta_e$. 
In other words, for every edge $e = \{k, \ell\}$ where $k < \ell$, the vector $v$ satisfies:
\[ v_k - v_\ell = \begin{cases}
1 & k = i, \ell = j \\
0 & \text{otherwise}
\end{cases}  \,.\]
From this, we see that $v_j = v_i + 1$ for our edge of interest.

Now, suppose towards contradiction that the edge $\{i, j\}$ is not a global bridge. 
Then,  there exists some sequence of nodes $\{i, m_1, \dots, m_p, j \}$ which connects $i$ and $j$ through a sequence of edges $\{i, m_1\}, \dots, \{m_p, j\}$
However, the existence of this path would imply $v_i = v_{m_1} = \dots = v_{m_p} = v_j$, contradicting the equality $v_j = v_i + 1$ above.
Therefore, if $I_e^g =1$, then $e = \{i, j\}$ is a global bridge.

$(\Leftarrow)$ Suppose $e = \{i, j\}$ is a global bridge.  Then, we can show $\delta_e \in \textrm{im} (B_1^\top)$. 
Specifically, consider the vector $v$ which is $1$ for all nodes connected to $i$ when the edge $e$ is removed, and equal to $0$ for all nodes connected to $j$ when edge $e$ is removed from the graph (note that this is possible because the removal of $e$ disconnects the graph).  
We see that $B_1^\top v= \delta_e$, and hence $\delta_e \in \textrm{im} (B_1^\top)$.  
As a result, it must be the case that $\delta_e^g = \delta_e$, and therefore $I_e^g =1$ as desired.

\paragraph{Proof of Proposition \ref{prop:curl}} Given a simplicial complex $\mathcal{X}$ and and edge $e$, we wish to prove $I_e^c > 0$ if and only if $e$ is a part of at least one filled triangle, i.e. there exists $t \in \mathcal{X}$ such that $|t| = 3$ and $e \subset t$.

Because $I_e^c = \|\delta_e^c\|$,  we first characterize $\delta_e^c$.
Since $\delta_e^c$ is the projection of $\delta_e$ onto $B_2$, we can write:
\[ \delta_e^c = B_2 (B_2^\top B_2)^\dagger B_2^\top \delta_e \,, \]
which clearly has a term of the form $B_2^\top$.

$(\Rightarrow)$ If $I_e^c > 0$, then it must be the case $\delta_e^c \neq 0$. Hence, $B_2 (B_2^\top B_2)^\dagger B_2^
\top \delta_e \neq 0$ and it must be the case $B_2^\top \delta_e \neq 0$.
Therefore, there must be an entry in $B_2^\top$ for which the corresponding row for edge $e$ has a non-zero entry, i.e. there exists some $t$ such that $e \subset t$.

$(\Leftarrow)$ If there exists some $t$ such that $e \subset t$, then $B_2^\top \delta_e \neq 0$, as the entry of $B_2^\top \delta_e$ corresponding to the triangle $t$ must have a non-zero entry. 
Moreover, because $B_2^\top \delta_e \in \textrm{im}(B_2^\top)$, we must have $B_2 (B_2^\top B_2)^\dagger B_2^
\top \delta_e \neq 0$.  Hence,  $\delta_e^c \neq 0$ and $I_e^c > 0$, proving the claim.

\subsubsection{Theoretical Results on Edge PageRank}

Theorem \ref{thm:hodge_informal} shows that the Edge PageRank measure can be understood as a function of a weighted Hodge Decomposition.
In this section, we first define the weighted Hodge Decomposition, which we use throughout the rest of the section. 
\begin{definition}[(Weighted) Hodge Decomposition \cite{schaub2018random}]\label{def:weighted_hodge}
    For a vector $v \in \mathbb{R}^{|E|}$, the weighted Hodge Decomposition of $v$ is a set of 3 vectors $v^{wg}, v^{wc}$, and $v^{wh}$ such that $v = v^{wg} + v^{wc} + v^{wh}$.
    Specifically, $v^{wg}$ is the projection of $v$ onto $D_2 B_1^\top$ under the metric defined in the inner product space $\langle x, y \rangle_{D_2^{-1}} = x D_2^{-1} y$, $v^{wc}$ is the projection of $v$ onto $B_2$ in the same inner product space, and $v^{wh}$ is defined as $v - v^{wg} - v^{wc}.$
\end{definition}
In the above definition, $D_2$ is the same normalization matrix considered in Eq. \eqref{eq:normalized_hodge_laplacian}.
Moreover, we note that the indicator flow of a particular edge can similarly be rewritten as a function of this weighted Hodge Decomposition

\paragraph{Proof of Theorem \ref{thm:hodge_informal}}

In this section, we first formally state Theorem \ref{thm:hodge_informal} and then provide a proof of the claim using the definition of the weighted Hodge Decomposition.
We formally restate Theorem \ref{thm:hodge_informal} as follows:

\begin{theorem} \label{thm:hodge}
    For a simplicial complex $\mathcal{X}$, an edge $e$, and a parameter $\beta > 2$, the Edge PageRank vector $\pi_e$ can be written:
    \[ \pi_e =\frac{(\beta - 2)}{\beta} \left[ \delta_e^{wh} +  \left(I + \frac{D_2 B_1^\top D_1^{-1} B_1}{\beta}\right)^{-1} \delta_e^{wg} + 
    \left(I + \frac{B_{2} D_3 B_{2}^\top D_{2}^{-1}}{\beta}\right)^{-1} \delta_e^{wc} \right] \,,\]
    where we recall that $\delta_e \in \mathbb{R}^{|E|}$ is the indicator vector which is $1$ at the index corresponding to edge $e$ and $0$ elsewhere, that  $B_1, B_2, D_1, D_2, $ and $D_3$ are as in Definition \ref{def:normalized_laplacian} of the main text, and that $\delta_e^{wg}$, $\delta_e^{wc}$, and $\delta_e^{wh}$ represent the weighted gradient, curl, and harmonic components of the indicator vector as noted in the definition above.
\end{theorem}
\begin{proof}
    Let us consider the following dynamical system with state $\pi_e^t \in \mathbb{R}^{|E|}$, where $t$ denotes a discrete time index starting at $t = 0$.
    As an initial condition, let $\pi_e^0 = 0$, and let the dynamics be as follows:
    \[ \pi_e^{t+1} = -\frac{\mathcal{L}_1}{\beta} \pi_e^t + \frac{(\beta - 2)\delta_e}{\beta} \,.\]
    We will show the fixed point of this dynamic system is exactly equal to the Edge PageRank vector, and can also be written in the form of the theorem.
    
    The fixed point of this dynamical system satisfies:
    \[ \pi_e^* = -\frac{\mathcal{L}_1}{\beta} \pi_e^* + \frac{(\beta - 2)\delta_e}{\beta} \implies (\beta I +  \mathcal{L}_1) \pi_e^* = (\beta - 2) \delta_e \,,\]
    and hence the solution to the system is exactly the Edge PageRank vector in Definition 3 of the main text.
    Moreover, we see that $\pi_e^*$ can also be written as:
    \begin{align*}
        \pi_e^* = \lim_{t \rightarrow \infty} \pi_e^t = \sum_{j = 0}^\infty \left(-\frac{\mathcal{L}_1}{\beta} \right)^j \frac{(\beta - 2) \delta_e}{\beta} \,.
    \end{align*}
    We note that by the definition of the weighted Hodge Decomposition, $\delta_e^{wg} = D_2 B_1^\top v_e$ for some vector $v_e$, $\delta_e^{wc} = B_2 t_e$ for some vector $t_e$, and $\mathcal{L}_1 \delta_e^{wh} = 0$, where $\delta_e^{wg} + \delta_e^{wc} + \delta_e^{wh} = \delta_e$.
    Hence,
    \begin{align*}
        \pi_e^* &= \sum_{j = 0}^\infty \left(-\frac{\mathcal{L}_1}{\beta} \right)^j \frac{(\beta - 2) (\delta_e^{wg} + \delta_e^{wc} + \delta_e^{wh})}{\beta} \\
        &= \frac{(\beta - 2)\delta_e^{wh} }{\beta} +  \sum_{j = 0}^\infty \left(-\frac{\mathcal{L}_1}{\beta} \right)^j \frac{(\beta - 2) (D_2 B_1^\top v_e + B_2 t_e)}{\beta} \,.
    \end{align*}
    Further, note the following fact about $\mathcal{L}_1$ for $j \geq 1$:
    \begin{align*}
        (\mathcal{L}_1)^j = (D_2 B_1^\top D_1^{-1} B_1)^j + (B_2 D_3 B_2^\top D_2^{-1})^j \,,
    \end{align*}
    which follows from the definition of $\mathcal{L}_1$ and the fact that $B_1 B_2 = 0$.
    Therefore, we must have:
    \begin{align*}
         \pi_e^* &= \frac{(\beta - 2)\delta_e^h }{\beta}  \\
         &\qquad + \sum_{j = 0}^\infty \left(-\frac{D_2 B_1^\top D_1^{-1} B_1}{\beta} \right)^j \frac{(\beta - 2) (D_2 B_1^\top v_e + B_2 t_e)}{\beta} \\
         &\qquad + \sum_{j = 0}^\infty \left(-\frac{B_2 D_3 B_2^\top D_2^{-1}}{\beta} \right)^j \frac{(\beta - 2) (D_2 B_1^\top v_e + B_2 t_e)}{\beta} \\
         &= \frac{(\beta - 2) \delta_e^h}{\beta}  \\
         &\qquad + \sum_{j = 0}^{\infty} \left( - \frac{D_2 B_1^\top D_1^{-1} B_1}{\beta}\right)^j \frac{(\beta - 2)D_2 B_1^\top v_e}{\beta}  \\
        &\qquad + \sum_{j = 0}^{\infty} \left( - \frac{B_{2} D_3 B_{2}^\top D_{2}^{-1}}{\beta}\right)^j \frac{(\beta - 2) B_2 t_e}{\beta} \\
        &= \frac{(\beta - 2) \delta_e^h}{\beta} +  \sum_{j = 0}^{\infty} \left( - \frac{D_2 B_1^\top D_1^{-1} B_1}{\beta}\right)^j \frac{(\beta - 2) \delta_e^{wg}}{\beta}  + 
    \sum_{j = 0}^{\infty} \left( - \frac{B_{2} D_3 B_{2}^\top D_{2}^{-1}}{\beta}\right)^j \frac{(\beta - 2) \delta_e^{wc}}{\beta}\,,
    \end{align*}
    where the penultimate simplification follows from the fact that $B_1 B_2 = 0$, and the final simplification follows from the definitions of $\delta_e^{wg}$ and $\delta_e^{wc}$.
    Finally, we note that if $\beta > 2$, then each of the infinite sums are guaranteed to converge, due to the definitions of the normalization matrices (this follows from the interpretation of the normalized $1$-Laplacian as a random walk in a lifted space \cite{schaub2018random}, which implies the spectral radius of $\mathcal{L}_1$ is at most 2).
    This results in the following equivalent representation of $\pi_e^*$:
    \begin{equation} \label{eq:inv}
        \pi_e^* = \frac{(\beta - 2)}{\beta} \delta_e^{wh} +  \left(I + \frac{D_2 B_1^\top D_1^{-1} B_1}{\beta}\right)^{-1} \frac{(\beta - 2)\delta_e^{wg}}{\beta}  + 
    \left(I + \frac{B_{2} D_3 B_{2}^\top D_{2}^{-1}}{\beta}\right)^{-1}\frac{(\beta - 2)\delta_e^{wc} }{\beta}  
    \end{equation}
    Hence, because $\pi_e^*$ is equal to the Edge PageRank vector $\pi_e$, and also can be represented as in Eq. \eqref{eq:inv}, the Theorem follows. 
\end{proof}

\end{document}